\documentclass[aps,pre,reprint,floatfix,superscriptaddress]{revtex4-1}
\usepackage{graphicx}
\usepackage[latin1]{inputenc}
\usepackage{amsmath}
\usepackage{amsfonts}
\usepackage{amssymb}
\usepackage{xcolor}

\newcommand{\blue}{\textcolor{blue}}

\begin{document}

\title{Driven toroidal helix as a generalization of Kapitzas pendulum}

\date{\today}

\author{J.~F.~Gloy}
\affiliation{Zentrum f\"ur Optische Quantentechnologien, Fachbereich Physik, Universit\"at Hamburg, Luruper Chaussee 149, 22761 Hamburg Germany} 
\author{A.~Siemens}
\email{asiemens@physnet.uni-hamburg.de}
\affiliation{Zentrum f\"ur Optische Quantentechnologien, Fachbereich Physik, Universit\"at Hamburg, Luruper Chaussee 149, 22761 Hamburg Germany} 
\author{P.~Schmelcher}
\email{pschmelc@physnet.uni-hamburg.de}
\affiliation{Zentrum f\"ur Optische Quantentechnologien, Fachbereich Physik, Universit\"at Hamburg, Luruper Chaussee 149, 22761 Hamburg Germany}
\affiliation{Hamburg Center for Ultrafast Imaging, Universit\"at Hamburg, Luruper Chaussee 149, 22761 Hamburg Germany}

\begin{abstract}

\noindent We explore a model system consisting of a particle confined to move along a toroidal helix while being exposed to a static potential as well as a driving force due to a harmonically oscillating electric field. 
It is shown that in the limit of a vanishing helix radius the governing equations of motion coincide with those of the well-known Kapitza pendulum - a classical pendulum with oscillating pivot - implying that the driven toroidal helix represents a corresponding generalization. 
It is shown that the two dominant static fixed points present in the Kapitza pendulum are also present for a finite helix radius. 
The dependence of the stability of these two fixed points on the helix radius, the driving amplitude, and the static potential are analyzed both analytically and numerically. 
Additionally, the most prominent deviations of the driven helix from Kapitzas pendulum with respect to the resulting phase space are investigated and analyzed in some detail. 
These effects include an unusual transition to chaos and an effective directed transport due to the simultaneous presence of multiple chaotic phase space regions. 

\end{abstract}

\maketitle

\section{Introduction}
\label{sec:Introduction}

Helical shapes are naturally occurring in nature, arising e.g. through hydrogen bonds in alpha-helix segments of proteins, or in molecules such as DNA and alpha-keratin. 
Furthermore, helical structures can emerge through long-range order in self-organizing systems on cylindrical surfaces \cite{kohlstedt2007,vernizzi2009,srebnik2011} or can be artificially created by rolling up thin sheets into cylinders \cite{lau2006,rodriguez1995,shaikjee2012,schmidt2001}. 
Helical structures can also appear in trapping potentials induced by light fields around optical fibers \cite{reitz2012,bhattacharya2007} which can be loaded with neutral cold atoms. 
An advantage of the helical shape is the increased stability with regards to deformations \cite{marko1994,kornyshev2007} making helical nano-structures desirable for future applications e.g. in nano circuits. 

Besides occurring in nature, helical systems of charged particles have recently been explored in the literature thereby demonstrating a number of intriguing effects emerging due to the geometry, such as interactions that oscillate with the (parametrized) distance along the helix \cite{schmelcher2011}. 
These effects have been studied in lattice systems with long-range hopping \cite{stockhofe2015,yu2001}, as well as in more fundamental models of classical charges moving on helices \cite{zampetaki2013,zampetaki2015,zampetaki2015a,zampetaki2017,plettenberg2017,zampetaki2018,siemens2020,siemens2021}. 
In such model systems, it has been demonstrated that based on the oscillating effective interactions already static setups become very complex since particles are able to localize into irregular lattice-like structures \cite{zampetaki2015,zampetaki2018} exhibiting a plethora of possible equilibrium configurations \cite{schmelcher2011,siemens2020}. 
By varying the helix geometry it is possible to tune a variety of effects, such as scattering of bound states at local defects \cite{zampetaki2013}, band structure inversion and degeneracies \cite{zampetaki2015,zampetaki2015a} or unusual pinned to sliding transitions \cite{zampetaki2017} in crystalline configurations on a toroidal helix. 

Inspired by the demonstrated richness of effects of charged particles on a helix, we explore here a system consisting of a single particle confined to a toroidal helix in the presence of an oscillating driving field and a static potential. 
In a previous study \cite{siemens2021} the corresponding phase space in the absence of the static potential and the related directed transport have been investigated. 
Here, we build upon these results and explore the effects of an additional spatially oscillating static potential. 
We show that the governing equations map to the equations for the Kapitza pendulum \cite{kapitza1951} in the limit of a vanishing helix radius. 
For a non-vanishing helix radius, a dynamical behavior beyond that of Kapitza's pendulum emerges. 
Our main results include a stability analysis of two major fixed points corresponding to the two major fixed points in Kapitza's pendulum. 
We derive and analyze some of the most prominent novel dynamical phases arising in the phase space of our driven helical particle system. 

This work is structured as follows: 
In section \ref{sec:DrivenTorHel}, we explain our setup and derive the underlying equations of motion. 
We show that in the limit of a vanishing helix radius the equations of motion simplify to those of Kapitzas pendulum. 
Therefore the main features of Kapitzas pendulum are briefly summarized in section \ref{sec:KapitzaLimit}. 
The main results are provided in sections \ref{sec:StabilityAnalysis} and \ref{sec:PhaseSpaceAnalysis} addressing the driven helix away from the Kapitza limit. 
In section \ref{sec:StabilityAnalysis} the influence of a finite helix radius on the stability of the two fixed points of Kapitzas pendulum are analyzed both analytically and numerically. 
In Section \ref{sec:PhaseSpaceAnalysis} major dynamical effects emerging for a finite helix radius are investigated. 
A discussion and outlook are presented in section \ref{sec:DiscussionAndOutlook}.

\section{Driven toroidal helix}
\label{sec:DrivenTorHel}

We consider a charged particle with charge $q$ confined to frictionlessly move on a geometry of the shape of a toroidal helix (see Fig. \ref{figure1}(a)). 
Additionally, the particle is subject to a static potential and driven by a harmonically oscillating electric field.
The confining geometry is parametrized as follows
\begin{equation}\label{eq:1}
\textbf{r}(u):=\left(
\begin{array}{c}
\left(R+r\cos(u)\right)\cos(u/M) \\
\left(R+r\cos(u)\right)\sin(u/M) \\
r\sin(u)
\end{array}
\right),
u\in [0,2\pi M]
\end{equation}
where $M$, $R$, $r$ are the number of helix windings, the radius of the torus, and the radius of the helix respectively. %, and $u$ characterizes the position on the helix. 
The parametrized position $u$ on the helix can be interpreted as an angle. 
If $u$ changes by $2\pi$, the particle moves by exactly one helical winding. 
The static potential $V(u)$ at each position $\textbf{r}(u)$ is defined as 
\begin{equation}\label{eq:2}
V_S(u)=V_0 \cos\left(\dfrac{u}{M}\right) % cos(u/M)
\end{equation}
The potential created by the periodic driving electric field $\textbf{E}(t)$ is modeled according to the corresponding Stark term
%TODO E-FIELD DIRECTION: WE ONLY CONSIDER X!!!
\begin{equation}\label{eq:3}
V_E(u,t)=q\textbf{E}(t)\cdot\textbf{r}(u)=q\cos(\omega t)\textbf{E}_0\cdot\textbf{r}(u)
\end{equation}
We consider a sinusoidally oscillating electric field with a polarization in the torus plane (x-direction). 
With this, the potential energy induced by the driving field becomes
\begin{equation}\label{eq:4}
V_E(u,t)=qE_0 \left(R + r cos(u)\right) cos(u/M)\cos(\omega t)
\end{equation}
This potential consists of two parts: one depending on the torus radius $R$ and one depending on the helix radius $r$. 
They will from now on be referred to as torus-induced potential (TIP) and winding-induced potential (WIP), respectively. 
An understanding of the potential experienced by the particle while moving along the helix can be gained from Fig. \ref{figure1}(b).
%-----------FIGURE EXPLANATION---------------------
The figure shows the energy due to the static potential (blue curve, compare Eq. \ref{eq:2}) and the energy due to the driving field at $t=0$ (orange and green curve for the TIP and WIP respectively, see Eq. (\ref{eq:4})) for a toroidal helix with $M=10$, $R=2.5$, $V_0=5$ and $r=0.8$. 
The total potential $V_{tot}(u,t)=V_E(u,t)+V_S(u)$ contains both the static potential $V_S(u)$ and the field potential $V_E(u,t)$. % is indicated by the pink dotted line. 
Due to the time dependence of the driving field the total potential energy is of course also time-dependent. 
Specifically, the shown TIP and WIP will oscillate with $\cos(\omega t)$, resulting in the total energies $V_{tot}(u,t=0)$ (pink dotted line in Fig. \ref{figure1}(b)) for a field aligned in positive x-direction, and  $V_{tot}(u,t=0.5\pi/\omega)$ (pink densely dotted line in Fig. \ref{figure1}(b)) half a driving period later when the field is aligned in the negative x-direction. 
The pink shaded area indicates the range of potential energies covered for each position $u$ during a driving period. 
An increase of $r$ will lead to an increase of the amplitude of the WIP. 
In the limit of $r\rightarrow0$ the WIP will vanish and the `fine structure' of $V_{tot}$ disappears. 
The number of extrema in the total potential energy can therefore be tuned by varying $r$. 
%---------END FIGURE EXPLANATION-------------------

The driven helix is then described by the following Lagrangian
\begin{equation}\label{eq:5}
\begin{array}{c}
\mathcal{L}=\dfrac{m}{2}\left(\dfrac{d\textbf{r}(u)}{dt}\right)^2-q\cos(\omega t)\textbf{E}_0\cdot\textbf{r}(u)+V_0 \cos\left(\dfrac{u}{M}\right) \\
=\frac{m}{2}\left(r^2 + a^2\left(R+r\cos(u)\right)^2\right)\dot{u}^2 \\
 -qE_0\left(R +r\cos(u)\right)\cos(\omega t)\cos(au)  - V_0 \cos(au)
\end{array}
\end{equation}
where $a=1/M$ is the inverse of the winding number. 
It is sensible to introduce the parameter $l(u)$ defined as 
\begin{equation}\label{eq:6}
l^2(u) := \frac{1}{a^2}\left(r^2 + a^2\left(R+r\cos\left(u\right)\right)^2\right)
\end{equation}
Using this expression the Lagrangian can be written as
\begin{equation}\label{eq:7}
\begin{array}{c}
\mathcal{L} = \frac{ma^2}{2}l^2(u)\dot{u}^2 \\ - \left( V_0 + qE_0\frac{\sqrt{l^2(u)a^2-r^2}}{a}\cos(\omega t)\right)\cos(au)

\end{array}
\end{equation}

The above Lagrangian efficiently accounts for the confining forces by only allowing positions along the helix $\textbf{r}(u)$. 
From this Lagrangian we obtain the following equation of motion
\iffalse
\begin{equation}\label{eq:8}
\begin{array}{c}
ma^2r\sin(u)\left(R+r\cos(u)\right)\dot{u}^2 /2 \\
+m\left[r^2+a^2\left(R+r\cos(u)\right)^2\right]\ddot{u} \\
-qE_0\cos(\omega t)\left[r\sin(u)\cos(au)+a\left(R+r\cos(u)\right)\sin(au)\right] \\
-V_0a\sin(au) = 0
\end{array}
\end{equation}
\fi
\begin{equation}\label{eq:8}
\begin{array}{c}
m\left[r^2+a^2\left(R+r\cos(u)\right)^2\right]\ddot{u} -V_0a\sin(au) \\
-qE_0\cos(\omega t)\left[r\sin(u)\cos(au)+a\left(R+r\cos(u)\right)\sin(au)\right] \\
+ma^2r\sin(u)\left(R+r\cos(u)\right)\dot{u}^2 /2 = 0
\end{array}
\end{equation}

Some of the parameters in the Lagrangian of Eq. (\ref{eq:7}) and of the equation of motion in Eq. (\ref{eq:8}) are redundant and can be `absorbed' by other parameters. 
The redundant parameters are: the driving frequency $\omega$, the torus radius $R$, the particle mass $m$, and charge $q$ of the particle. 
These quantities can without loss of generality be eliminated by rescaling the remaining relevant parameters as follows
\begin{equation}\label{eq:9}
\begin{array}{c}
\tilde{t} = t\frac{\omega}{2\pi}, \quad \tilde{r} = \frac{r}{R},  \quad \tilde{E} = \frac{4\pi^2qE}{mR\omega^2}, \quad \tilde{V} = \frac{4\pi^2V}{mR^2\omega^2} 
\end{array}
\end{equation}

In the limit of $r\rightarrow0$, we get $l^2(u)=R^2$, and the Lagrangian from Eq. (\ref{eq:7}) becomes the Lagrangian of Kapitzas pendulum \cite{kapitza1951}:
\begin{equation}
\mathcal{L}_{K} = \frac{m}{2}a^2R^2\dot{u}^2 + \left( V_0 + qE_0 R\cos(\omega t)\right)\cos(au)
\label{eq:10}
\end{equation}
The equivalence between Kapitzas pendulum and the toroidal helix in the limit of $r\rightarrow0$ is further indicated in Fig. \ref{figure1}(c) and its inset. 
The driving electric field and static potential along the toroidal helix are, respectively, equivalent to the oscillating pivot and the gravitation potential in Kapitzas pendulum.

\section{The Kapitza pendulum limit}
\label{sec:KapitzaLimit}

\begin{figure}
\includegraphics[width=\columnwidth]{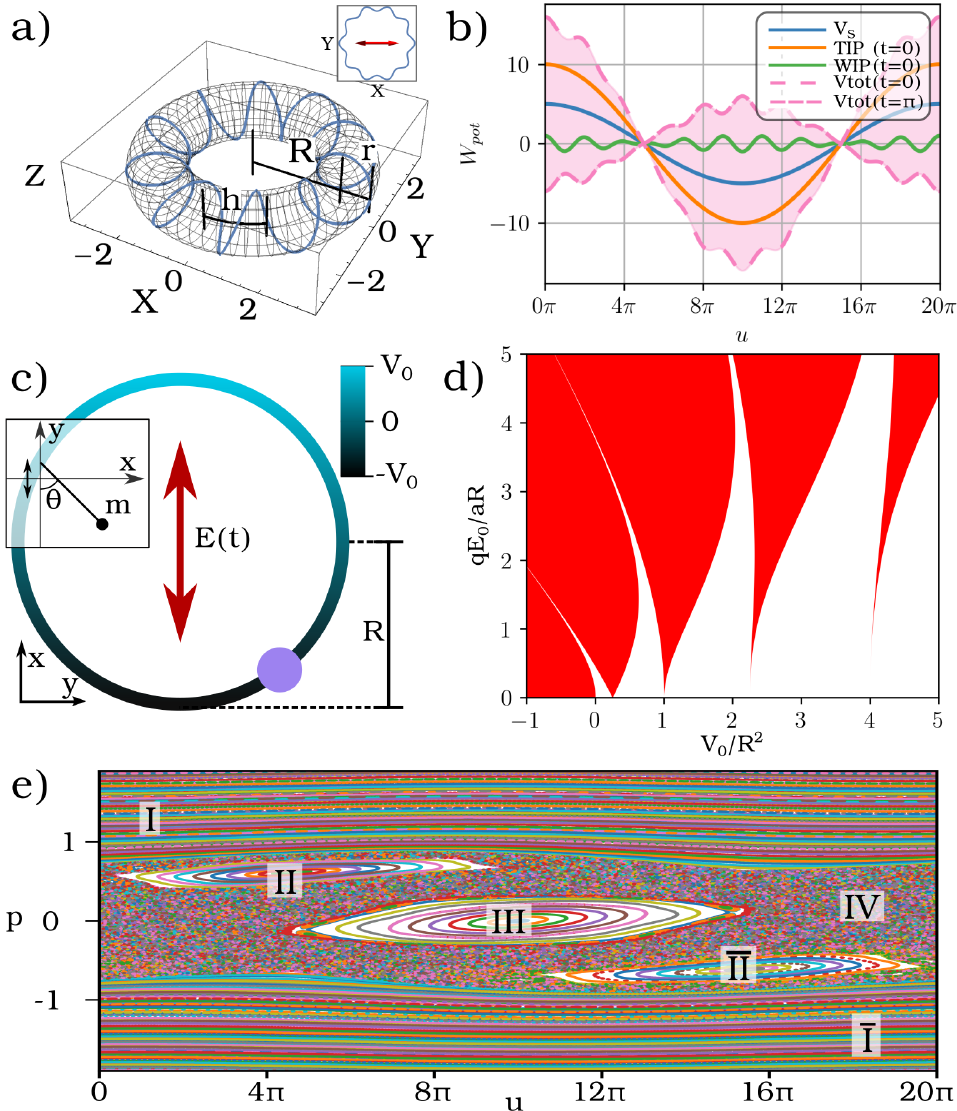}
\caption{\label{figure1} (a) A 3D sketch of the torus and the toroidal helix with the parametric function $\textbf{r}(u)$, for $M=10$, $r=0.8$, and $R=2.5$. The inset in the top right visualizes the direction of the driving electric field. (b) The potential energy created by the driving electric field $E(t)$ (TIP, orange and WIP, green) and the static potential $V(u)$ (blue) shown for a toroidal helix with $M=10$, $R=2.5$, $V_0=5$, and a helix radius of $r=0.1$. (c) Visualization of the Kapitza limit $r\rightarrow0$. The toroidal helix becomes a circle in the xy-plane. The potential energy induced by the static potential is indicated by the color. For comparison, a schematic of Kapitzas pendulum is shown in the inset on the left. (d) Visualization of the Ince-Strutt diagram highlighting the regions where the two major fixed points of Kapitzas pendulum are stable (white) or unstable (red). (e) Poincare Surface of Section (PSOS) in the Kapitza limit $r\rightarrow0$ for $V_0=5$ and $E_0=3$. The most prominent types of trajectories  are shown: (I) rotators that are not significantly affected by the driving, (II) trajectories circling in phase around the ring with the driving, (III) bounded trajectories centered at the minimum of the static potential, and (IV) chaotic trajectories. Trajectories marked $\overline{\text{II}}$ circle around the ring in the opposite direction as those trajectories marked II. }
\end{figure}

To be self-contained, we briefly demonstrate the main features of our system that are already known from the Kapitza pendulum. 
The Kapitza pendulum is a classical pendulum with oscillating pivot as depicted in the inset of Fig. \ref{figure1}(c). 
One of the most interesting aspects of Kapitzas pendulum regards the fixed points in the underlying equations of motion. 
In addition to the expected fixed point where the pendulum is in its potential minimum (corresponding to $u=M\pi$), Kapitzas pendulum can have another stable fixed point in the upper position (corresponding to $u=0$).
This second fixed point is stabilized due to the driving forces from the oscillating pivot. 
In the Kapitza limit of $r\rightarrow0$ the equation of motion shown in Eq. (\ref{eq:8}) simplifies to %becomes
\begin{equation}\label{eq:11}
ma^2 R^2\ddot{u}= \left[V_0a + qE_0\cos(\omega t)aR\right]\sin(au)
\end{equation}
From Eq. (\ref{eq:11}) the two fixed points at $u=0$ and $u=M\pi$ - respectively corresponding to the Kapitza pendulum in the upper and lower position - can be easily identified. 
The stability of these fixed points can be determined by linearizing Eq. (\ref{eq:11}) around these two fixed points. 
This results in the following equation
\begin{equation}\label{eq:12}
mR^2\ddot{u}=u\left(\pm V_0 +qE_0 R\cos(\omega t)\right)
\end{equation}
where in case of the fixed point at $u=0$ we obtain a positive sign of the first summand, and a negative sign in case of the fixed point at $u=M\pi$.
Equation (\ref{eq:12}) is also known as the Mathieu equation (compare Eq. (\ref{eq:14}) below). 
The parameter regions for which the Mathieu equation has periodic bounded solutions can be determined from the Ince-Strutt diagram \cite{butikov2018} shown in Fig. \ref{figure1}(d). 
In this diagram the white areas mark regions where periodic solutions of Eq. (\ref{eq:12}) exist i.e. where the fixed point is stable, whereas in the red regions, no bounded solutions exist i.e. the fixed point is unstable. 
As can be seen from Eq. (\ref{eq:12}), positive values on the ($V_0/R^2$)-axis of Fig. \ref{figure1}(d) describe the stability of the fixed point at $u=0$, whereas negative values describe the stability of the fixed point at $u=M\pi$.

The below-given discussions in Sections \ref{sec:StabilityAnalysis} and \ref{sec:PhaseSpaceAnalysis} feature an analysis of the phase space for $r>0$ to understand the dynamics for a wide range of initial conditions. 
To better contextualize these results, the most prominent types of trajectories in the Kapitza limit are now discussed. 
Since our phase space is made up of three parameters (position $u$, momentum $p$ and time $t$) we can use a Poincar\'e surface of section (PSOS) - specifically a stroboscopic map - to visualize the phase space in a two dimensional stroboscopic $u(p)$ mapping. 
Note, that our momentum $p$ refers to the canonical momentum given by
\begin{equation}\label{eq:13}
p=\dfrac{du/dt}{m\left( r^2 + ((R+r\cos(u))/M)^2 \right)}
\end{equation}
A general overview of the most prominent possible types of trajectories in the Kapitza limit is given in Fig. \ref{figure1}(e). 
For large enough momentum there will always be trajectories corresponding to a fast motion which is not significantly affected by the driving. 
They are marked (I) in the figure (or $\overline{\text{I}}$ for trajectories moving in the opposite direction.
Islands of regular motion around the two fixed points at $u=0$ and $u=M\pi$ and are marked (III). 
Additionally, it is possible to stabilize (quasi-)periodic trajectories circling around the torus in phase with the driving field. 
This type of motion occurs in the regions marked (II) and ($\overline{\text{II}}$) in Fig. \ref{figure1}(e). 
Chaotic trajectories (marked (IV)) will in general be present for all $E_0>0$. 
Through variation of $E_0$ and $V_0$ it is possible to tune the presence of the trajectories of type (II), (III), and (IV). 
All of these trajectories are also encountered for arbitrary $r>0$ - albeit for different parameter combinations than for $r=0$.

\section{Stability analysis}
\label{sec:StabilityAnalysis}

\begin{figure*}
\includegraphics[width=\linewidth]{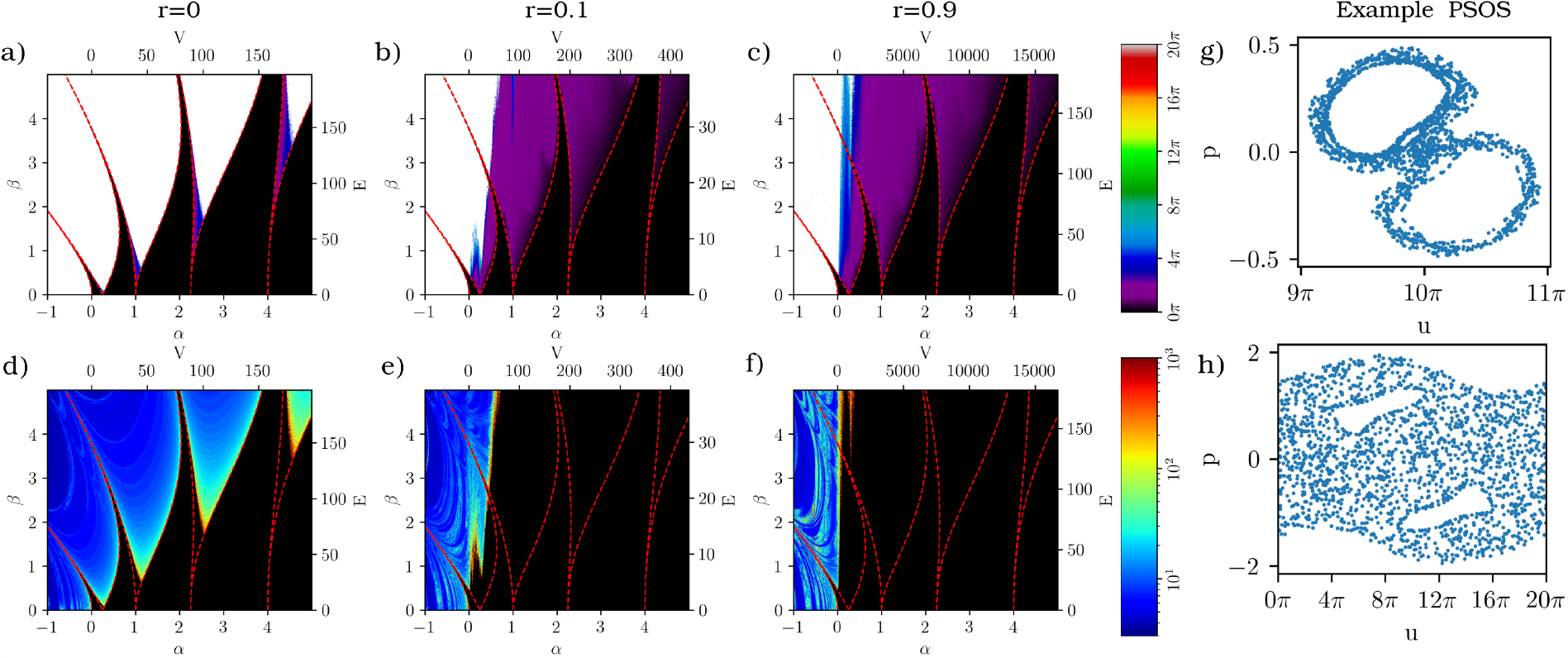}
\caption{\label{figure2} Numerical stability analysis. (a-c) Maximal distance in phase space between the fixed point (at $u=M\pi$ for $\alpha>0$ and at $u=0$ for $\alpha<0$) and a trajectory starting at a distance of $10^{-8}$ from this point after a simulation time of $1000$ driving periods. White color indicates that the particle moves at least once around the torus. The parameter regions where our analytical calculations based on Eqs. (\ref{eq:14}) to (\ref{eq:16}) predict the fixed points to be stable are marked by the dotted red lines. (d-f) The time needed for the particle to move once around the torus. The used trajectories are the same as in (a-c). Again, the corresponding Ince-Strutt diagram is indicated by the dotted red lines. (g-h) Example PSOS for trajectories from the purple and white regions. }
\end{figure*}

We will now consider the general case of a finite helix radius $r>0$ and investigate how the helix radius influences the stability of the two fixed points discussed in section \ref{sec:KapitzaLimit}. 
The persistence of these fixed points in the generalized setup can be directly verified by inserting the initial conditions $\dot{u}=0$ and $u=0$ (or $u=m\pi$ for the second fixed point) into the general equations of motion given by Eq. (\ref{eq:8}). 
In addition to an analytical stability analysis through linearization of the equations of motion in the vicinity of the fixed point, we investigate the particle dynamics close to the fixed point via numerical simulations. 
Note that from now on, all calculations are performed using the scaling introduced in Eq. (\ref{eq:9}).
We start with the analytical considerations and linearize the general equations of motion in Eq. (\ref{eq:8}) around the two fixed points. 
Similar to the Kapitza limit, the resulting approximate equations of motion are described by the Mathieu equation, except that this time the coefficients of the Mathieu equation additionally depend on the helix radius. 
The general Mathieu equation is given by 
\begin{equation}\label{eq:14}
\ddot{u}+\left(\alpha-\beta\cos(\tau)\right)u=0 
\end{equation}
For the first fixed point at $u=0$ the parameters $\alpha$ and $\beta$ are given by 
\begin{equation}\label{eq:15}
\begin{aligned}
\alpha_1 &= -\frac{V_0 a^2}{4\pi^2( r^2 + a^2(1 +r)^2)}  \\
\beta_1 &= \frac{ E_0 \left(a^2(1 + r) +r \right)}{4\pi^2( r^2 + a^2(1 +r)^2)}
\end{aligned}
\end{equation}
For the second fixed point at $u=M\pi$ we have 
\begin{equation}\label{eq:16}
\begin{aligned}
\alpha_2 &= \frac{V_0 a^2}{4\pi^2(r^2 + a^2(1 + (-1)^M r)^2)} \\
\beta_2 &= \frac{ E_0 \left(a^2(1 + (-1)^M  r) + (-1)^M r \right)}{4\pi^2(r^2 + a^2(1 + (-1)^M r)^2)}.
\end{aligned}
\end{equation}
The factor $(-1)^M$ in Eq. (\ref{eq:16}) accounts for the difference in the potential energy at $u=M\pi$ between setups with even and odd winding numbers. 
In the following, all shown data are for an even winding number $M=10$.
For even $M$ we get $\alpha_1=-\alpha_2$ and $\beta_1=\beta_2$ and we can therefore visualize the stability of both fixed points in the same diagram. 
Different choices (i.e. odd values) of $M$ will change the parameters $\alpha$ and $\beta$ but to the best of our knowledge do not lead to significantly different behavior or dynamics.

Using Eqs. (\ref{eq:14}) to (\ref{eq:16}), we can establish and analyze the Ince-Strutt diagram to determine the parameter sets for which the two fixed points of our driven helix are stable. 
This is illustrated in Fig. \ref{figure2} for several values of the helix radius $r$. 
The boundaries of the analytically obtained stability `tongues' (i.e. regions where the fixed points are predicted to be stable) obtained from Eqs. (\ref{eq:14}) to (\ref{eq:16}) are marked by red dotted lines. 
The colors in Fig. \ref{figure2} visualize the results of the numerical stability analysis and provide insight into the dynamics in the immediate vicinity of the fixed points. 
These numerical results are obtained by calculating the trajectory of a particle starting within an $\epsilon$ environment of the fixed points. 
If the fixed point is stable, the resulting motion is (quasi-) periodic; in case it is unstable, the particle will explore a significant region of the phase space. 
More specifically, we use the initial condition of $(u,p)=(M\pi+10^{-8},0)$ and simulated the dynamics for $1000$ driving periods. 
For each trajectory, the maximal phase space distance of the trajectory to the fixed point is determined. 
These results are shown in Figs. \ref{figure2}(a-c), where each pixel corresponds to a distance obtained from a single trajectory. 
In total, $675000$ trajectories were simulated for each of the sub-figures Fig. \ref{figure2} (a-c). 
The black areas indicate that the particle stays in the immediate vicinity of the fixed point, whereas the white color shows that the particle moves at least once around the torus. 
The agreement with the analytically determined stability diagrams can be clearly seen in Fig. \ref{figure2}. 
However, an increase of the helix radius $r$ leads to a significant change of the dynamics of unbounded trajectories for the fixed point at $u=M\pi$ (i.e. positive values of $V_0$ in the figure). 
Increasing $r$, increases the size of regions where the particle moves a significant distance away from the fixed point but does not explore the complete phase space (i.e. the purple and blue regions in the figure). 
In the white regions of the figure, the unstable fixed point is (usually) part of the chaotic sea, allowing the particle to explore the entire toroidal helix. 
A PSOS for a corresponding example trajectory can be seen in Fig. \ref{figure2}(h). 
An example PSOS for a trajectory from the purple and blue regions is shown in Fig. \ref{figure2}(g). 
The dynamics in the blue and purple regions of the figures will be described in more detail in section \ref{sec:PhaseSpaceAnalysis}.

In contrast to the fixed point at $u=M\pi$, judging from Figs. \ref{figure2}(a-c) the behavior outside of the stability tongue for the fixed point at $u=0$ (negative values of $V_0$ in the figure) seems to be hardly affected by changes of $r$. 
One intuitive explanation for this is that $V_S(u)$ has a maximum at $u=0$ and at infinitesimal distances from this point the particle will experience a force away from the fixed point, thereby preventing the existence of trajectories similar to the one shown in Fig. \ref{figure2}(g).

To provide insight into the trajectories in the white regions,we determine the time needed until a distance of $2\pi M$ is reached for the first time. 
The corresponding results are shown in Figs. \ref{figure2}(d-f). 
We observe that for increasing $r$ the transition from (quasi-)periodic to chaotic trajectories in the vicinity of the stability-tongue borders changes from a (relatively) smooth transition for $r=0$ to a rather abrupt transition for large $r$.

%TODO: EXPLAIN DYNAMICS IN ALL CASES (STABLE,UNSTABLE,CHAOTIC)...

%TODO: FOR V SMALLER THAN E WE WILL ALWAYS HAVE A POTENTIAL MINIMUM BUT NOT ALWAYS STABLE MOTION!

\section{Phase-space analysis}
\label{sec:PhaseSpaceAnalysis}

\begin{figure}
\includegraphics[width=\columnwidth]{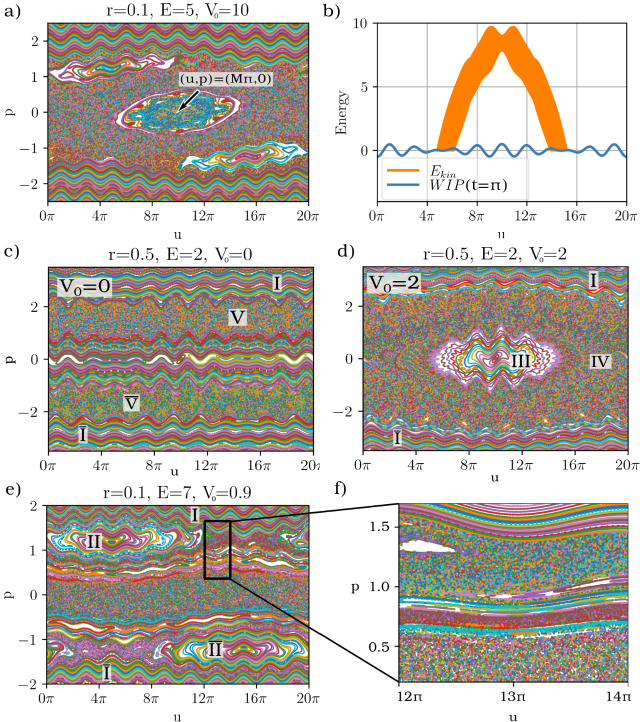}
%\caption{\label{figure3} (a) Poincare surface of sections (PSOS) for $r=0.1$, $E_0=5$ and $V_0=10$. Trajectories close to the fixed point at $(u,p)=(M\pi,0)$ (indicated by the arrow) become chaotic while (quasi-) periodic trajectories with greater phase space distances from the fixed point prevail. (b) Potential energy due to the WIP, as well as the kinetic energy of one of the (quasi-) periodic trajectories that separates the two chaotic regions shown in (a). (c) PSOS for $r=0.5$, $E_0=2$ and $V_0=0$ (left panel), $V_0=2$ (right panel). The two separated chaotic seas in the left panel for $V_0=0$ become connected when $V_0$ is of similar order as $E_0$. (d) PSOS for $r=0.1$, $E_0=7$ and $V_0=0.9$ showing the presence of three distinct chaotic regions. The inset highlights the upper- and part of the center chaotic regions. 
\caption{\label{figure3} 
(a) Poincare surface of sections (PSOS) for $r=0.1$, $E_0=5$ and $V_0=10$. Trajectories close to the fixed point at $(u,p)=(M\pi,0)$ (indicated by the arrow) become chaotic while (quasi-) periodic trajectories with greater phase space distances from the fixed point prevail. 
(b) Potential energy due to the WIP, as well as the range of kinetic energy values taken by one of the (quasi-) periodic trajectories that separates the two chaotic regions shown in (a). 
(c-d) PSOS for $r=0.5$, $E_0=2$ and $V_0=0$ (c), $V_0=2$ (d). The two separated chaotic seas marked V and $\overline{\text{V}}$ in (c) become connected when $V_0$ is of similar order as $E_0$. 
(e-f) PSOS for $r=0.1$, $E_0=7$ and $V_0=0.9$ showing the presence of three distinct chaotic regions. The upper chaotic region is highlighted in (f). %The inset highlights the upper part of the center chaotic regions. A close-up highlighting the presence of the additional chaotic regions. 
}
\end{figure}

In addition to the modifications of the stability of the two fixed points of Kapitzas pendulum, the driven helix also exhibits various dynamical `phases' that appear only for a non-vanishing radius $r>0$. 
In this section the most significant of these features are described and analyzed. 

One interesting characteristic concerns the unusual mechanism by which the dynamics in the vicinity of the fixed point at $u=M\pi$ transitions from (quasi-)periodic to chaotic motion. 
It might be natural to expect that the break-up of invariant tori happens first for those trajectories with larger phase-space distance to the fixed point when the system is exposed to a perturbation.  
However, in contrast to this expectation, we observe that trajectories close to the fixed point become chaotic - resulting in a chaotic phase space region that is centered around an unstable fixed point and separated from the `main' chaotic sea by a region with (quasi-) periodic trajectories i.e. a regular island. 
The size of this chaotic region can be tuned by varying the helix radius $r$. 
The result are chaotic trajectories with a tunable motional `amplitude' (i.e. tunable maximal distance from the fixed point) around the fixed point. 
This effect is demonstrated in the PSOS in Fig. \ref{figure3}(a).  
In the figure one can clearly identify the (quasi-)periodic trajectories and regular regions separating the chaotic trajectories that are trapped around the fixed point at $(u,p)=(M\pi,0)$ from those chaotic trajectories that can explore the entire toroidal helix.
For small parameter regions, the simultaneous presence of multiple `bands' of (quasi-) periodic trajectories centered around the (unstable) fixed point at $(u,p)=(M\pi,0)$, but separated from one another by chaotic phase space regions, could be observed.

The reason for this peculiar transition to chaos can be elucidated by the changes in the potential landscape for increasing $r$. 
For $r\ll R$, the radius dependent oscillations (the WIP, with a period of $2\pi$) of the potential can be treated as a perturbation to the $r\rightarrow0$ limit. 
This perturbation will be largest at the global extrema of the potential at $u=0$ and $u=M\pi$ and will vanish at $u=M\pi/2$ and $u=3M\pi/2$ (see Fig. \ref{figure1}(b)). 
This heavy oscillatory character of the time-dependent potential landscape can induce chaotic motion - provided the particle moves slow enough to be affected. 
From a comparison with Fig. \ref{figure1}(e), it can be seen that this is more likely for trajectories oscillating closer to the fixed point and less likely with increasing phase space distance of the trajectory from the fixed point. 
Consequently, trajectories closer to the fixed point (i.e. closer to the global minimum of the potential landscape) will be stronger affected by this perturbation and will become chaotic for smaller values of $r$ than their more distant counterparts. 
These arguments are supported by Fig. \ref{figure3}(b) which shows the WIP-potential energy at $t=\pi$, together with the set of kinetic energy  values $\text{\{} E_{kin}(u(t)) \, | \, 0<t<2000\pi \text{\}}$ taken by the particle during a representative (quasi-)periodic example trajectory confining the chaotic phase space region around the fixed point. 
It can be seen that the kinetic energy is for the most part much larger than the perturbation by the WIP. 
Only close to those points where the WIP vanishes does the kinetic energy also become comparatively small.

Another interesting effect that is absent in the Kapitza limit concerns the emergence of chaos in the regime of weak driving forces. 
In the regime of small driving amplitudes two separate chaotic phase space regions, that are arranged symmetrically with respect to a point inversion symmetry through the point $(u,p)=(M\pi,0)$, can appear - one consisting of trajectories with only positive momenta, the other consisting of trajectories with only negative momenta (see regions marked V and $\overline{\text{V}}$ in Fig. \ref{figure3}(c)). 
They are similar to the trajectories marked (II) and ($\overline{\text{II}}$) in Fig. \ref{figure1}(e) in the sense that they also correspond to motion around the torus with strictly positive or negative momentum. 
However, instead of moving once around the torus during each driving period, these trajectories are chaotic and move on average by one helix winding during each driving period. 
The average velocity in these trajectories is therefore slower by a factor of $1/M$ compared to the average velocity of the Type-II (and $\overline{\text{II}}$) trajectories. 
These trajectories appear only in the case of a finite helix radius $r$. 
The (quasi-) periodic trajectories separating the two chaotic regions correspond mostly to very slow (quasi-) periodic motion of the particle around the torus and in some cases to (quasi-) periodic oscillations of the particle within one helix winding. 
The origin and mechanism of this effect has previously been explored in the absence of the static potential \cite{siemens2021}. 
When the static potential is added, the two chaotic regions will persist while $V_0\ll E_0$. 
However, when $V_0$ is increased, the chaotic regions also increase and will fuse when $V_0$ is of similar order of magnitude as $E_0$, thereby resulting in a phase space similar to the one shown in Fig. \ref{figure3}(d). 
In the figure, $V_0$ is sufficiently large, such that all slowly moving (quasi-) periodic trajectories will be part of the regular island around the fixed point at $(u,p)=(M\pi,0)$ (marked III in Fig. \ref{figure3}(d)) and none of the separating trajectories persist. 

Another interesting effect concerns the influence of a finite helix radius on the trajectories moving around the torus in phase with the driving (see regular islands marked II and $\overline{\text{II}}$ in Fig. \ref{figure1}(e)). 
As shown in Fig. \ref{figure3}(e-f), chaotic regions separated from the `main' chaotic region which is centered around $p=0$ can appear around these regular islands. 
The dynamics in the chaotic regions that surround the regular islands marked II and $\overline{\text{II}}$ in Fig. \ref{figure3}(e) correspond to motion where the particle moves around the torus (on average) in phase with the driving frequency.  
A necessary condition for this effect to occur is, that the driving amplitude is small enough, such that the chaotic sea centered around $p=0$ does not surround the corresponding two regular islands. 
Analogous to the effect shown in Figs. \ref{figure3}(a-b), these chaotic regions are caused by perturbations of the trajectories due to the WIP. 
One difference to this previously discussed effect is that the regular islands marked II and $\overline{\text{II}}$ are respectively located at the positions $u=M\pi/2$ and $u=3M\pi/2$ where the WIP vanishes. 
The perturbation is consequently stronger for trajectories with larger phase space distances from the fixed point.

\section{Summary and Discussion}
\label{sec:DiscussionAndOutlook}

We have demonstrated that the dynamics of a charged particle confined to a toroidal helix while being exposed to a static potential and a driving electric field represent a generalization of Kapitzas pendulum in the sense that in the limit of a vanishing helix radius their equations of motion coincide. 
We discuss the effects of a finite helix radius while focusing on two different aspects: the stability of the two prominent fixed points of Kapitzas pendulum, and the impact of a nonzero helix radius on the structure of the phase space and the corresponding dynamics. 
For a finite helix radius the dynamics in the linearized neighborhood of the main fixed points can be approximated by a Mathieu equation with modified parameter values. 
From this, the general stability of both fixed points for different driving amplitudes $E_0$, static potential amplitudes $V_0$, and helix radii $r$ have been determined. 
Our analytical results agree with those of corresponding numerical simulations. 
The latter show that the dynamics in the extended neighborhood of the fixed point at $u=M\pi$ can change significantly for increasing $r$, whereas for the fixed point at $u=0$ no such changes could be observed. 
Specifically, the change in dynamics can be directly observed in the phase space, where for an increasing helix radius the fixed point at $(u,p)=(M\pi,0)$ can undergo an unusual transition to chaos. 
Additionally, two other prominent dynamical `phases' that only appear for finite helix radii have been discovered. 
These `phases' are characterized by the presence of multiple separate chaotic seas in the phase space. 
Especially notable is that the presence of multiple chaotic seas allows for chaotic particle trajectories with non-zero average velocity (i.e. directed transport), even though the spatio-temporal symmetries that are usually associated with a vanishing directed transport (here $(u,p,t)\rightarrow (-u Mod(2M\pi),-p,t)$ and $(u,p,t)\rightarrow (u,-p,-t)$)  are not broken by the driving field. 
Notable are also the (quasi-) periodic trajectories separating the two chaotic seas for small driving amplitudes and finite $r$, since they correspond to regular (directed) motion with very low momentum around the torus. 
%TODO: EXPLAIN COUNTERINTUITIVITY OF THESE TRAJECTORIES ...

The observed dynamics in our driven helix to be seen as a generalized Kapitza pendulum is a direct consequence of the additional winding induced potential appearing in the corresponding equations of motion. 
Some of the described effects are even occurring in parameter regimes where the WIP can be treated as a perturbation to Kapitzas pendulum. 
Therefore, an educated guess would be that other periodic position-dependent small amplitude perturbations of Kapitzas pendulum will result in a dynamic similar to the one observed here. 
Consequently, we expect that many of the described effects can be found e.g. in a mechanical Kapitza pendulum with position-dependent length. 
%A direct experimental implementation of the Lagrangian in Eq. \ref{eq:7} could be realized in the context of ultracold atoms. 
%There, the possibility of trapping neutral atoms along a helical path in light fields around optical fibers \cite{reitz2012} has recently been demonstrated. 
%However, the realization of the driving forces in such a system may be a challenge. 

%\begin{acknowledgments}
%A.S. thanks ... for fruitful discussions.  
%\end{acknowledgments}

%\bibliographystyle{apsrev4-2}
\bibliographystyle{apsrev4-1}

%%%\bibliography{txtest.bib}
\bibliography{txtest}
%\bibliography{HelicalDipoles}

\end{document}